\documentclass[fleqn,usenatbib]{mnras}
\usepackage{newtxtext,newtxmath}

\usepackage[T1]{fontenc}

\DeclareRobustCommand{\VAN}[3]{#2}
\let\VANthebibliography\thebibliography
\def\thebibliography{\DeclareRobustCommand{\VAN}[3]{##3}\VANthebibliography}

\usepackage{svg}
\usepackage{graphicx}	
\usepackage{amsmath}	

\usepackage{tikz,xcolor,hyperref}

\definecolor{lime}{HTML}{A6CE39}
\DeclareRobustCommand{\orcidicon}{%
	\begin{tikzpicture}
	\draw[lime, fill=lime] (0,0) 
	circle [radius=0.16] 
	node[white] {{\fontfamily{qag}\selectfont \tiny ID}};
	\draw[white, fill=white] (-0.0625,0.095) 
	circle [radius=0.007];
	\end{tikzpicture}
	\hspace{-2mm}
}

\foreach \x in {A, ..., Z}{%
	\expandafter\xdef\csname orcid\x\endcsname{\noexpand\href{https://orcid.org/\csname orcidauthor\x\endcsname}{\noexpand\orcidicon}}
}

\title[Tidal Orbital Decay in Hot Jupiters]{Searching for Tidal Orbital Decay in Hot Jupiters}

\author[Alvarado III]{Efrain Alvarado III\orcidA{}$^{1}$\thanks{E-mail: 
efrain.alvarado.iii@berkeley.edu},
Kate B. Bostow\orcidD{}$^{1}$,
Kishore C. Patra\orcidC{}$^{1,2}$ \thanks{kcpatra@berkeley.edu},
Cooper H. Jacobus\orcidB{}$^{1,3}$,
\newauthor
Raphael A. Baer-Way$^{1,4}$,
Connor F. Jennings$^{1}$,
Neil R. Pichay$^{1}$,
Asia A. deGraw$^{1}$,
Edgar P. Vidal\orcidJ{}$^{1}$,
\newauthor
Vidhi Chander$^{1}$,
Ivan A. Altunin$^{1}$,
Victoria M. Brendel$^{1}$,
Kingsley E. Ehrich$^{1,5}$,
James D. Sunseri\orcidI{}$^{1,6}$,
\newauthor
Michael B. May$^{1,7}$,
Druv H. Punjabi$^{1}$,
Eli A. Gendreau-Distler\orcidH{}$^{1}$,
Sophia Risin $^{1,3}$,
Thomas G. Brink\orcidF{}$^{1,8}$,
\newauthor
WeiKang Zheng\orcidG{}$^{1,9}$,
and Alexei V. Filippenko\orcidE{}$^{1}$
\\
$^{1}$Department of Astronomy, University of California, Berkeley, CA 94720-3411, USA\\
$^{2}$Nagaraj-Noll-Otellini Graduate Fellow in Astronomy \\
$^{3}$Computational Cosmology Center, Lawrence Berkeley National Laboratory, Berkeley, CA 94720, USA\\ 
$^{4}$Department of Astronomy, University of Virginia, Charlottesville, VA 22904, USA \\
$^{5}$Department of Astronomy, The University of Florida, Gainesville, FL 32611, USA \\
$^{6}$Department of Astrophysical Sciences, Princeton University, 4 Ivy Lane, Princeton, NJ 08540, USA \\
$^{7}$Department of Nuclear Engineering, University of Tennessee, Knoxville, TN 37996, USA \\
$^{8}$Draper-Wood-Robertson Specialist in Astronomy \\
$^{9}$Bengier-Winslow-Eustace Specialist in Astronomy \\
}

\date{Accepted 2024 August 30. Received 2024 August 29; in original form 2024 February 13}

\pubyear{2024}

\begin{document}
\label{firstpage}
\pagerange{\pageref{firstpage}--\pageref{lastpage}}
\maketitle

\begin{abstract}
We study transits of several ``hot Jupiter'' systems -- including WASP-12\,b, WASP-43\,b, WASP-103\,b, HAT-P-23\,b, KELT-16\,b, WD\,1856+534\,b, and WTS-2\,b -- with the goal of detecting tidal orbital decay and extending the baselines of transit times. We find no evidence of orbital decay in any of the observed systems except for that of the extensively studied WASP-12\,b. Although the orbit of WASP-12\,b is unequivocally decaying, we find no evidence for acceleration of said orbital decay, with measured $\ddot{P} = (-7 \pm 8) \times 10^{-14} \rm ~s^{-1}$, against the expected acceleration decay of $\ddot{P} \approx -10^{-23} \rm ~s^{-1}$.
In the case of WD 1856+534\,b, there is a tentative detection of orbital growth with $\dot{P} = (5.0 \pm 1.5) \times 10^{-10}$. While statistically significant, we err on the side of caution and wait for longer follow-up observations to consider the measured $\dot{P}$ real. For most systems, we provide a 95\%-confidence lower limit on the tidal quality factor, $Q_\star'$. The possibility of detecting orbital decay in hot Jupiters via long-term radial velocity (RV) measurements is also explored. We find that $\sim 1 \rm ~m~s^{-1}$ precision in RVs will be required to detect orbital decay of WASP-12\,b with only 3~yr of observations. Currently available RV measurements and precision are unable to detect orbital decay in any of the systems studied here.

\end{abstract}

\begin{keywords}
planets and satellites: dynamical evolution and stability -- planet–star interaction -- techniques: photometric
\end{keywords}



\section{Introduction}

The discovery of 51 Pegasi b -- a large, Jupiter-like planet orbiting its host star with an orbital period of $\sim 4$ days -- marked the first time an exoplanet was found orbiting a main-sequence star \citep{1995Natur.378..355M}.At the time, it was widely believed that Jupiter-sized planets formed far from their host stars as in our own Solar System, but this discovery challenged this previously held notion about formation of planets and their migration (see \citealt{2018ARA&A..56..175D} for a review).  Since then, over 400 such planets have been discovered\footnote{https://exoplanetarchive.ipac.caltech.edu/}. Some defining characteristics of these exoplanets, known as ``hot Jupiters,'' are that they have a short orbital period ($< 10$ days) and they orbit close to their host star ($\lesssim 0.1$~au). 

Even now, the origins and fates of such exoplanet systems are uncertain. Theoretically, it is expected that most hot Jupiters
will gradually spiral into their host stars as a result of dissipation of energy via tidal interactions, and transfer of angular momentum from the planet's orbit to the star's spin \citep{2009ApJ...692L...9L, 1996ApJ...470.1187R, 2003ApJ...596.1327S}. The efficiency with which tidal energy is dissipated in the stellar interior is described by the tidal quality factor ($Q_\star'$), which is related to the timescale of orbital decay. However, this timescale has proven difficult to calculate theoretically, because of our incomplete understanding of the physical processes by which the
energy of tidal oscillations is ultimately converted into heat
\citep{2014ARA&A..52..171O}. 

To date, WASP-12b is the only exoplanet confirmed to undergo orbital decay with a measured period derivative of $-29$  $\pm$  2 ms yr$^{-1}$ \citep{2020ApJ...888L...5Y,2020AJ....159..150P,2016A&A...588L...6M}. Transit observations of WASP-12\,b have therefore been invaluable in efforts to better understand the mechanisms of tidal dissipation in hot Jupiter systems \citep{2017ApJ...849L..11W,2019MNRAS.482.1872B,2021ApJ...918...16M}. 
More recently, hints of orbital decay have been detected in the Kepler-1658\,b system with a rate of $\dot{P} = 131\substack{+20 \\ -22}$ ms yr$^{-1}$ \citep{2022ApJ...941L..31V}. Observing transits and timing them over a long temporal baseline can help detect any changes in the dynamics of the star-planet system. For example, we can examine mass loss, which can expand or contract the orbit of the planet \citep{2015ApJ...813..101V,2016CeMDA.126..227J}, or apsidal precession of the planet's orbit \citep{2002ApJ...564.1019M}, or secular changes in the orbit via interactions with a distant third body \citep{2019AJ....157..217B}.

The goal of this paper is to add more transit times and thus increase the temporal baseline of observations for several hot Jupiters, specifically WASP-12\,b, WASP-43\,b, WASP-103\,b, KELT-16\,b, HAT-P-23\,b, WD 1856+534\,b, and WTS-2\,b. These systems are a subset of hot Jupiters most amenable to detection of orbital decay \citep{2020AJ....159..150P}. In addition to adding more transit times, we aim to search for evidence of tidal orbital decay within these systems and provide updated constraints on the tidal quality factor, $Q_\star'$. We also explore the possibility of detecting orbital decay with long-term monitoring of radial velocities (RVs) of these systems. 

The paper is organised as follows. Section \ref{New Times} describes the observations of new transit times for WASP-12\,b, WASP-43\,b, WASP-103\,b HAT-P-23\,b, KELT-16\,b, WD 1856+534\,b, and WTS-2\,b.  The methods and procedures for detecting orbital decay via transit and RV measurements are discussed in Section \ref{Analysis}. Section \ref{Results} presents our results, and we summarise our conclusions in Section \ref{Conclusions}.

\section{New Transit Times} \label{New Times}

Transits of WASP-12\,b, WASP-43\,b, WASP-103\,b, HAT-P-23\,b, KELT-16\,b, WD 1856+534\,b, and WTS-2\,b were observed with the 1\,m Nickel telescope at Lick Observatory on Mt. Hamilton, CA, USA. The telescope is equipped with a Loral $2048 \times 2048$ pixel CCD and has a field of view of $\sim 6.3^{\prime}$ on each side. We utilised $2 \times 2$ binning, resulting in a scale of $0.184^{\prime \prime}$ pixel$^{-1}$ for all images. Images were primarily taken with the Johnson $R$ filter, with the exception of the dates of observation of 2022 March 23 (UTC dates are used throughout this paper) for WASP-103 and 2022 April 2 for WD\,1856+534, which were both obtained with the $V$ filter. Additionally, we adjusted the exposure-time settings to compensate for varying weather conditions, with the goal of increasing the signal-to-noise ratio (S/N). The log of observations is provided in Table \ref{Nights}.

Each raw image was bias-frame subtracted and flat-field corrected using AstroImageJ (AIJ; \citealt{Collins_2017})\footnote{AIJ is an extension of ImageJ augmented to be an astronomy-specific image display environment with tools for astronomy-specific image calibration and data reduction.}. We used the Barycentric Julian Date in Barycentric Dynamical Time ($\mathrm{BJD}_{\mathrm{TDB}}$) timing standard for all observations. The mid-exposure time of each image at the observatory's local time was converted to $\mathrm{BJD}_{\mathrm{TDB}}$ using the time utility code of \cite{Eastman_2010}. We would like to stress the importance of adopting mid-exposure time of the image as the timestamp for a flux measurement rather than the beginning or the end of the exposure. Otherwise, different authors may inadvertently introduce systematic shifts in their transit timing analyses of up to half of the exposure duration -- a significant amount of error when attempting to detect small changes in orbital period.  

We performed aperture photometry for each target star and 6--8 comparison stars. The relative flux of the target star was determined by summing the flux of each comparison star and then dividing it by the flux of the target star. Photometry was done for circular apertures of varying sizes and the aperture with the highest S/N was ultimately used in the analysis.  

The light curves were normalised to unity outside the transit. To model the observed light curve, we used the \cite{2002ApJ...580L.171M} transit model with the following free parameters: mid-transit time ($T_{\mathrm{mid}}$), a scaled ratio of stellar radius in terms of the orbital semimajor axis ($R_\star/a$), impact parameter ($b=a\cos(i)$/$R_\star$, where $i$ is the orbital inclination), and a planet-to-star radius ratio ($R_p/R_\star$). In addition to the transit model, we corrected for airmass-related extinction by allowing the relative flux to be a linear function of time, giving two additional parameters. Although airmass does not change linearly with time, applying a linear correction was sufficient for all our observations, eliminating the need for higher-order correction.

To set the transit timescale, we kept the orbital period fixed to the most recent value found in the literature (see \citealt{2010exop.book...55W}). Quadratic limb-darkening coefficients for each target were fixed to the values interpolated from the tables given by \cite{2011A&A...529A..75C}, using the codes of \cite{2013PASP..125...83E}. We used \texttt{emcee}, an affine invariant Markov Chain Monte Carlo (MCMC) ensemble sampler, to determine the posterior distributions for the transit parameters \citep{2013PASP..125..306F}.

\begin{table*}
    \caption{Observation log and the measured mid-transit times}
    \begin{tabular}{c c c c c c c c c}
    \hline
    \hline
     Target & Start Obs. & End Obs. & Filter & Exp. & Airmass ($X$) & Epoch & $T_{\mathrm{mid}}$ & Uncertainty \\
      & (UTC) & (UTC) & & (s) & & & $\mathrm{BJD}_{\mathrm{TDB}}$ (days) & (days) \\
     \hline
     WASP-12 & 2022 Feb 5 & 2022 Feb 5 & $R$ & 20 & 1.16-1.16 & 2387 & 2459615.72741 & 0.00086 \\
      & 03:14:41 & 08:00:03 & & & & & & \\
     WASP-12 & 2022 Feb 6 & 2022 Feb 6 & $R$ & 20 & 1.01-1.66 & 2388 & 2459616.81979 & 0.00041 \\
      & 05:16:30 & 09:50:17 & & & & & & \\
      \hline
     WASP-43 & 2022 March 25 & 2022 March 25 & $R$ & 25 & 1.58-1.51 & 2754 & 2459663.75733 &  0.00022 \\
      & 04:55:06 & 07:09:28 & & & & & & \\
     WASP-43 & 2022 April 7 & 2022 April 7 & $R$ & 25 & 1.47-1.83 & 2770 & 2459676.77387 &  0.00030 \\
      & 05:04:36 & 07:41:30 & & & & & & \\
     \hline
     WASP-103 & 2022 March 23 & 2022 March 23 & $V$ & 24/40 & 1.65-1.17 & 2323 & 2459661.98526 & 0.00081 \\
     & 09:29:23 & 13:14:14 & & & & & & \\
     WASP-103 & 2022 April 5 & 2022 April 5 & $R$ & 24 & 1.67-1.17 & 2337 & 2459674.94535 & 0.00085 \\
     & 08:36:37 & 12:30:37 & & & & & & \\
     WASP-103 & 2022 April 18 & 2022 April 18 & $R$ & 24 &  1.92-1.18 & 2351 & 2459687.89960 & 0.00189 \\
     & 07:15:59 & 11:45:24 & & & & & & \\
     WASP-103 & 2022 June 7 & 2022 June 7 & $R$ & 24 & 1.17-1.69 & 2405 & 2459737.88136 & 0.00049 \\
     & 07:03:57 & 11:00:05 & & & & & & \\
     \hline
     HAT-P-23 & 2022 June 19 & 2022 June 19 & $R$ & 24/26/28/30 & 1.35-1.07 & 2679 & 2459749.90138 & 0.00030 \\
     & 07:52:58 & 11:11:45 & & & & & & \\
     HAT-P-23 & 2022 July 6 & 2022 July 6 & $R$ & 35 & 1.21-1.07 & 2693 & 2459766.88027 & 0.00043 \\
     & 07:30:13 & 10:08:02 & & & & & & \\
     HAT-P-23 & 2022 July 17 & 2022 July 17 & $R$ & 20 & 1.65-1.07 & 2702 & 2459777.79718 & 0.00038  \\
     & 05:09:06 & 09:00:19 & & & & & & \\
     \hline
     KELT-16 & 2022 June 29 & 2022 June 29 & $R$ & 24 & 1.58-1.00 & 1471 & 2459759.84606 & 0.00037 \\
     & 06:24:41 & 10:41:15 & & & & & & \\
     KELT-16 & 2022 July 30 & 2022 July 30 & $R$ & 24 & 1.10-1.07 & 1503 & 2459790.85362 & 0.00039 \\
     & 06:37:47 & 10:11:49 & & & & & & \\
     \hline
     WD 1856+534 & 2022 April 2 & 2022 April 2 & $V$ & 45 & 1.56-1.11 & 634 & 2459672.00859 & 0.00006 \\
      & 11:54:27 & 12:32:43 & & & & & & \\
     WD 1856+534 & 2022 June 3 & 2022 June 3 & $R$ & 120 & 1.04-1.06 & 678 & 2459733.95810 & 0.00008 \\
      & 10:37:02 & 11:18:00 & & & & & & \\
     \hline
     WTS-2 & 2022 March 25 & 2022 March 25 & $R$ & 120/150/180/240/270 & 1.85-1.16 & -50 & 2459663.98031 & 0.00046 \\
      & 10:35:59 & 13:00:26 & & & & & & \\
     WTS-2 & 2022 May 15 & 2022 May 15 & $R$ & 120 & 1.21-1.00 & 0 & 2459714.91443 & 0.00112 \\
      & 09:17:44 & 11:33:16 & & & & & & \\
     WTS-2 & 2022 July 2 & 2022 July 2 & $R$ & 120 & 1.32-1.00 & 47 & 2459762.79392 & 0.00034 \\
      & 05:36:53 & 08:29:49 & & & & & & \\
     WTS-2 & 2022 July 8 & 2022 July 8 & $R$ & 120 & 1.00-1.17 & 53 & 2459768.90638 & 0.00029 \\
      & 08:22:11 & 11:16:04 & & & & & & \\
     \hline
    \end{tabular}
\label{Nights}
\end{table*}

\section{Analysis} \label{Analysis}
\subsection{Timing Analysis}
Past transit times for our targets were gathered from the database of \cite{2022ApJS..259...62I}.
We fitted two models to the timing data. The first model assumes a constant orbital period,
\begin{equation}
    t(E) = t_0 + PE,
\label{const}
\end{equation}
where $E$ is the epoch number and $P$ is the orbital period. The second model assumes orbital decay,
\begin{equation}
    t(E) = t_0 + PE + \frac{1}{2}\frac{dP}{dE}E^2,
\label{decay}
\end{equation}
where $\frac{dP}{dE}$ is the rate of change of orbital period per orbit. From this, we can determine the rate of change of the orbital period, $\dot{P}$, as 
\begin{equation}
    \dot{P} = \frac{1}{P}\frac{dP}{dE}.
\end{equation}

In the case of WASP-12\,b, we also attempted to detect acceleration of the orbital decay. The acceleration model includes the third-order term in the Taylor expansion which can be written as

\begin{equation}
    t(E) = t_0 + PE + \frac{1}{2}\frac{dP}{dE}E^2 + \frac{1}{6}\frac{d^2P}{dE^2}E^3.
\label{acceleration}
\end{equation}
It follows that the acceleration of orbital decay can be calculated as 
\begin{equation}
    \ddot{P} = \frac{1}{P^2}\frac{d^2P}{dE^2}.
\end{equation}

We allowed $t_0$, $P$, $dP/dE$, and $d^2P/dE^2$ to be free parameters and used \texttt{emcee} to determine their posterior distributions. Table \ref{tab:Best Fits} lists the best-fit parameters for all models and Table \ref{tab:AllTimes} shows all transit times used in this work. 

Based on our best-fit value of $\dot{P}$ from the orbital decay model, we can determine $Q_\star'$ from
\begin{equation}
    Q_\star' = -\frac{27\pi}{2 \dot P}\left(\frac{M_p}{M_\star}\right)\left(\frac{R_\star}{a}\right)^5.
\label{eqn:Q_star_Pdot}
\end{equation}
This equation can be obtained by applying Kepler's third law to Eq.~20 of \cite{1966Icar....5..375G}, and is based on the traditional equilibrium tide model where the tidal bulge on the star lags behind the line joining the star and the planet by a constant angle. Equation \ref{eqn:Q_star_Pdot} can be differentiated and combined with Kepler's third law to get an analytic expression for $\ddot{P}$,
\begin{equation}\label{Pdotdot_analytic}
\begin{aligned}
\ddot{P} \approx 1.2 \times 10^{-24} ~\text{s}^{-1} \left(\frac{Q_\star'}{10^{6}} \right)^{-1}  
\left(\frac{M_{p}}{ M_{\text{J}}} \right)
\left(\frac{M_{\star}}{ M_{\odot}} \right)^{-8/3} 
\left(\frac{R_{\star}}{ R_{\odot}} \right)^{5} \\
\times 
\left(\frac{P}{\text{days}} \right)^{-13/3}
\left(\frac{\dot{P}}{10^{-9}} \right).
\end{aligned}
\end{equation}
Based on this expression, the expected acceleration of orbital decay in WASP-12\,b is $\ddot{P} \approx -3 \times 10^{-23} \rm~s^{-1}$.  

\begin{table*}
    \centering
    \caption{Best-Fit Parameters and $Q_\star'$ based on transit timing analysis}
    \begin{tabular}{c|c|c|c|c|c|c}     
         \hline
         \hline
         System & Parameters & Constant Period & Orbital Decay & Acceleration of Orbital Decay & $Q_\star'$\\
         \hline
         WASP-12 & $t_0$ ($\mathrm{BJD}_{\mathrm{TDB}}$) & 2457010.51204 (6) & 2457010.51298 (4) & 2457010.51298 (4) & (1.6$\pm$0.1)$\times$ $10^{5}$ \\
         & $P$ (days)& 1.09141892 (4) & 1.09141944 (2) & 1.09141948 (5) &   \\
         & $\dot{P}$ & & (-9.37 $\pm$ 0.33) $\times$ $10^{-10}$ & (-9.2$\pm$0.4)$\times 10^{-10}$ \\
         & $\ddot{P}$ (s$^{-1}$) & & &  (-7$\pm$8)$\times 10^{-14}$ \\
         \hline
         WASP-43 & $t_0$ ($\mathrm{BJD}_{\mathrm{TDB}}$) & 2457423.44971 (4) & 2457423.44973 (7) & &  $>3.9 \times 10^5$  \\
         & $P$ (days)& 0.81347405 (2) & 0.81347405 (2) \\
         & $\dot{P}$ &&  (-9 $\pm$ 51) $\times$ $10^{-12}$ \\
         \hline
         WASP-103 & $t_0$ ($\mathrm{BJD}_{\mathrm{TDB}}$) & 2457511.94449 (3) & 2457511.94449 (3) & & $> 1.18 \times 10^6$ \\
         & $P$ (days) & 0.92554540 (3) &  0.92554539 (6) \\
         & $\dot{P}$ & & (2.49 $\pm$ 9.94) $\times$ $10^{-11}$ \\
         \hline
         HAT-P-23 & $t_0$ ($\mathrm{BJD}_{\mathrm{TDB}}$) & 2456500.57792 (6) & 2456500.577914 (6) & & $> 9.4 \times 10^5$ \\
         & $P$ (days) & 1.21288648 (4) & 1.21288640 (7) \\
         & $\dot{P}$ & & (7.42 $\pm$ 5.77) $\times$ $10^{-11}$ \\
         \hline
         KELT-16 & $t_0$ ($\mathrm{BJD}_{\mathrm{TDB}}$) & 2458334.45807 (8) & 2458334.45816 (11) & & $> 2.2 \times 10^5$ \\
         & $P$ (days) & 0.96899282 (11) & 0.96899289 (12) \\
         & $\dot{P}$ & & (-3.94 $\pm$ 3.12) $\times$ $10^{-10}$ \\ 
         \hline
         WD 1856+534 & $t_0$ ($\mathrm{BJD}_{\mathrm{TDB}}$) & 2458779.375083 (2) & 2458779.375085 (2) & &  $>5.8 \times 10^{-5}$ \\
         & $P$ (days) & 1.40793922 (1) &  1.40793913 (2) \\
         & $\dot{P}$ & & (4.98 $\pm$ 1.54) $\times$ $10^{-10}$ \\
         \hline
         WTS-2 & $t_0$ ($\mathrm{BJD}_{\mathrm{TDB}})$ & 2459714.91497 (19) \\
         & $P$ (days) & 1.01870539 (12) \\
         \hline
    \end{tabular}\\
   {\raggedright \textbf{Note.} The numbers in parentheses are the $1\sigma$ uncertainties in the last few digits, e.g., 1 (4) means $1 \pm 4$, 0.44971 (4) means $0.44971 \pm 0.00004$, and 0.91497 (19) means $0.91497 \pm 0.00019$.
   Lower limits of $Q_\star'$ includes uncertainties from $M_p$, $M_\star$, $R_\star$, and $a$. WASP-12 is the only system where $Q_\star'$ is a proper measurement.
    \par}
    \label{tab:Best Fits}
\end{table*}

\begin{table*}
    \caption{Mid-transit times and Epochs}
    \centering
    \begin{tabular}{c|c|c|c|c}
        \hline
        \hline
         System & Epoch & $\mathrm{T}_{\mathrm{mid}}$ ($\mathrm{BJD}_{\mathrm{TDB}}$) & Uncertainty (days) & Reference \\ 
         \hline
         HAT-P-23 & 	-1540 &	2454632.73289 &	0.00073 &	\cite{2011ApJ...742..116B} \\
         HAT-P-23 &	-1469 &	2454718.84863 &	0.00048 &	\cite{2011ApJ...742..116B} \\
         HAT-P-23 &	-1465 &	2454723.69918 & 0.00039 &	\cite{2011ApJ...742..116B} \\
         HAT-P-23 &	-1423 &	2454774.64120 &	0.00056 &	\cite{2011ApJ...742..116B} \\
         HAT-P-23 &	-1285 &	2454942.01919 &	0.00088 &	\cite{2011ApJ...742..116B} \\
         \hline
    \end{tabular}\\
    {\raggedright \textbf{Note.} The full table is available online. See the Data Availability statement. 
    \par}
    \label{tab:AllTimes}
\end{table*}

\subsection{Radial-Velocity Analysis}

In this section, we examine the possibility of detecting orbital decay via long-term RV monitoring. We gathered RV measurements from past studies and identified systems that have more than 10 RV measurements, and span at least 2~yr of time. These criteria are admittedly arbitrary but nonetheless provide a reasonable starting point for selecting suitable targets. The only hot Jupiter systems in our study that satisfy these constraints are WASP-12, WASP-43, WASP-103, and HAT-P-23.

We fit two RV models to the data: Eq.~\ref{constant-model} (below), which describes RV as a function of time for an orbit with a constant period, and Eq.~\ref{decay-model} (below), which describes orbital decay of a planet. These models work under the assumption that these systems have circular orbits. This is a reasonable assumption, as orbits of most hot Jupiters are expected to circularise on timescales much shorter than the lifetimes of these planetary systems \cite{2014ARA&A..52..171O}. For an orbit with a constant period,

\begin{equation}
    \centering
    \mathrm{RV}(t) = K \sin \left (\frac{2 \pi}{P} t + \phi \right) + c,
    \label{constant-model}
\end{equation}
where $P$ is the orbital period, $\phi$ is the orbital phase, and $c$ is a constant that represents the residual barycentric velocity of the system in the radial direction. $K$ is the velocity semi-amplitude, which is given by
\begin{equation}
    \centering
    K = m_p\sin(i)(m_\star + m_p)^{-2/3}P^{-1/3},
\end{equation}
where $i$ is the orbital inclination; $m_p$ and $m_\star$ are the masses of the planet and the star, respectively. 

In the orbital decay model, we assume that the period decreases with constant $\dot{P}$. If the period is indeed decreasing, then the time interval between crests and troughs of the RV signal should decrease while $K$ should become larger over time. To account for this, we allow the period and $K$ to be a function of time:
\begin{equation}
    \mathrm{RV}(t) = K(t) \sin \left(\frac{2 \pi}{P(t)}t+\phi \right) + c,
    \label{decay-model}
\end{equation}
where
\begin{equation}
    K(t) = m_p {\rm sin}(i) (m_\star + m_p)^{-2/3} P(t)^{-1/3}
\end{equation}
and
\begin{equation}
    \centering
    P(t)=P_0+\frac{dP}{dt}t.
\end{equation}
$P_0$ is the orbital period at $t=0$.
 
\begin{figure*}
    \centering
    \includegraphics[width=\textwidth]{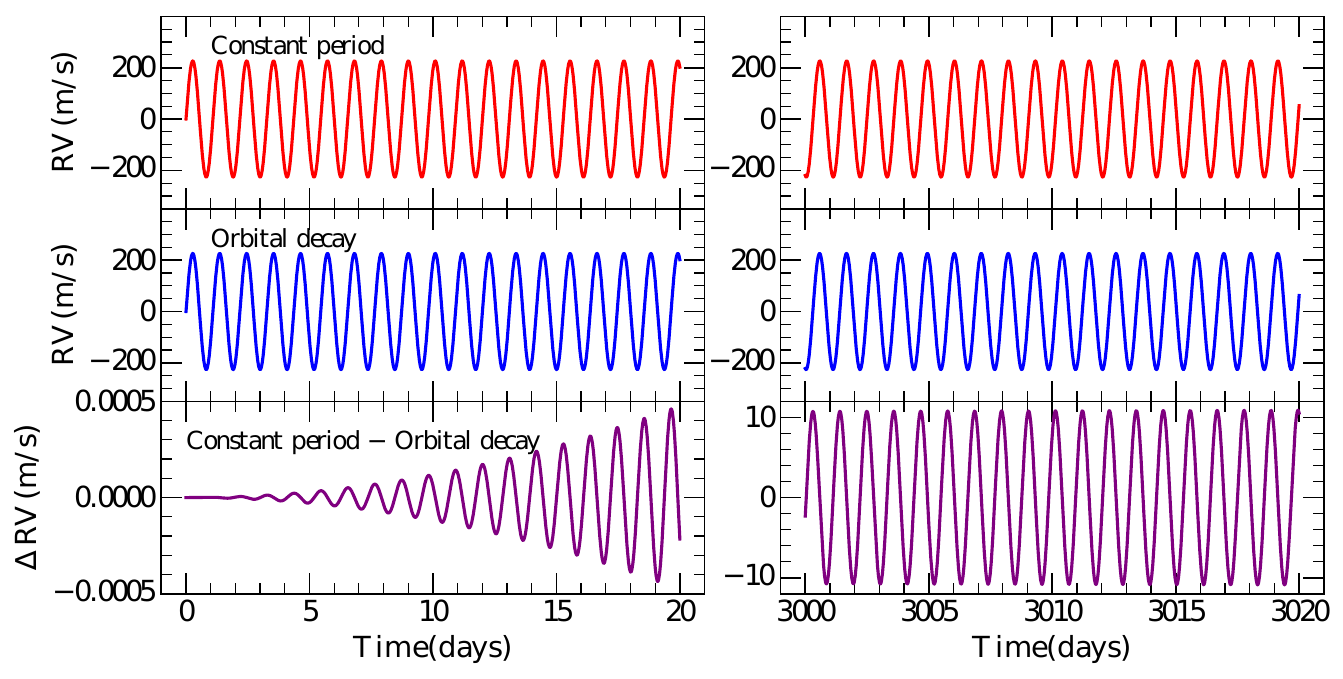}
    \caption{The RV signals for the constant-period and orbital-decay models. The left panels show the signal for the first 20 days, whereas the right panels show the signal after about 8 years. The bottom panel on both sides displays the difference between the two models. Note the divergence between the two models over time.}
    \label{fig:RV-Diff}
\end{figure*}

Figure \ref{fig:RV-Diff} shows the expected RV signal for the WASP-12\,b system for both the constant-period and the orbital-decay models. We see that the two models rapidly diverge and after about 8~yr; the models can differ by as much as $\sim 10 \rm ~m~s^{-1}$. This rapid divergence could potentially lead to the detection of orbital decay in as little as 3~yr of monitoring, if the precision on RV measurements approaches $\sim 1 \rm ~m~s^{-1}$. By comparison, the transit timing method required $\sim 10$~yr to detect orbital decay in WASP-12\,b. 

\begin{figure*}
    \centering
    \includegraphics[width=\textwidth]{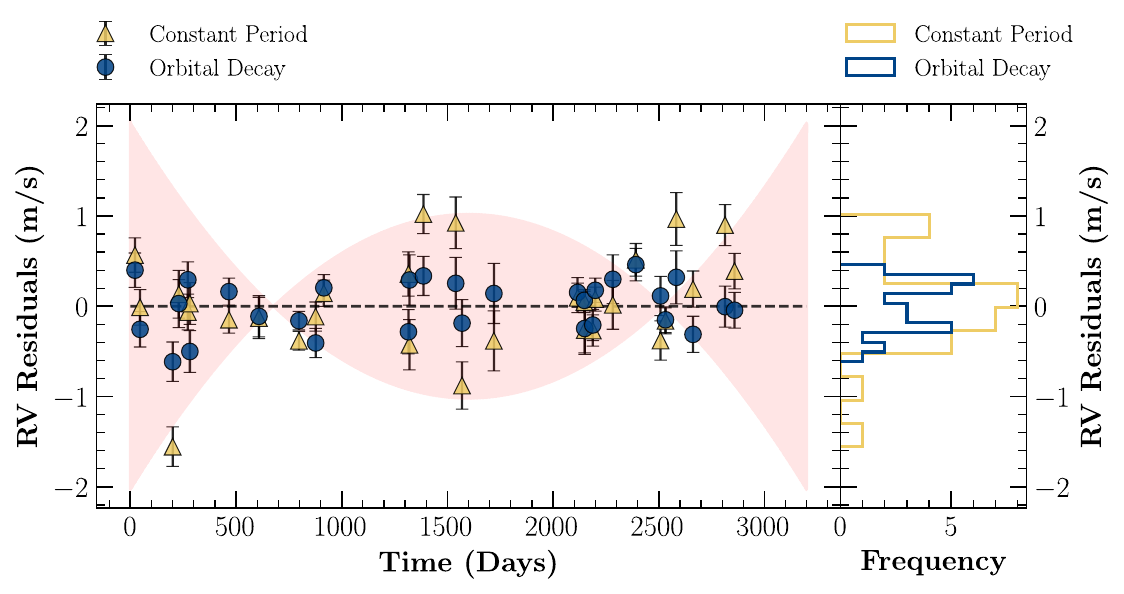}
    \caption{Left panel shows the residuals for the constant-period and the orbital-decay models fit to a simulated RV signal for a WASP-12\,b-like planet undergoing orbital decay. The semi-transparent red region shows where RV residuals are expected to lie when a constant-period model is fit to the RV signal of a decaying orbit. The right panel shows the distribution of the residuals between the two models.} 
    \label{MockData}
\end{figure*}

Next, we demonstrate how one may detect orbital decay in a WASP-12\,b-like planet with mock high-precision RV data. We randomly generated RV data points spanning over 3000 days following the currently known rate of orbital decay of WASP-12\,b. The uncertainty in RV measurements was sampled from the normal distribution $\sigma_{\rm RV} \sim \mathcal{N}(0.2, 0.0025 \rm ~m~s^{-1})$. We used \texttt{emcee} to fit the two aforementioned RV models to the mock data. The difference between the best-fit models and the simulated measurements is shown in Figure \ref{MockData} as RV residuals. As expected, the scatter in RV residuals is higher for the constant-period model since the mock data were generated assuming orbital decay. Consequently, when the orbital-decay model is fit, the scatter in residuals decreases. The shaded red region shows where RV residuals are expected to lie when a constant-period model is fit to the RV signal of a system that is intrinsically undergoing orbital decay. The red region is analogous to the quadratic transit-timing residual seen in WASP-12\,b. Figure \ref{MockData} demonstrates that for sufficiently high RV precision, it is possible, in principle, to detect orbital decay with long-term RV monitoring. 

\begin{figure}
    \centering
    \includegraphics[width=\columnwidth]{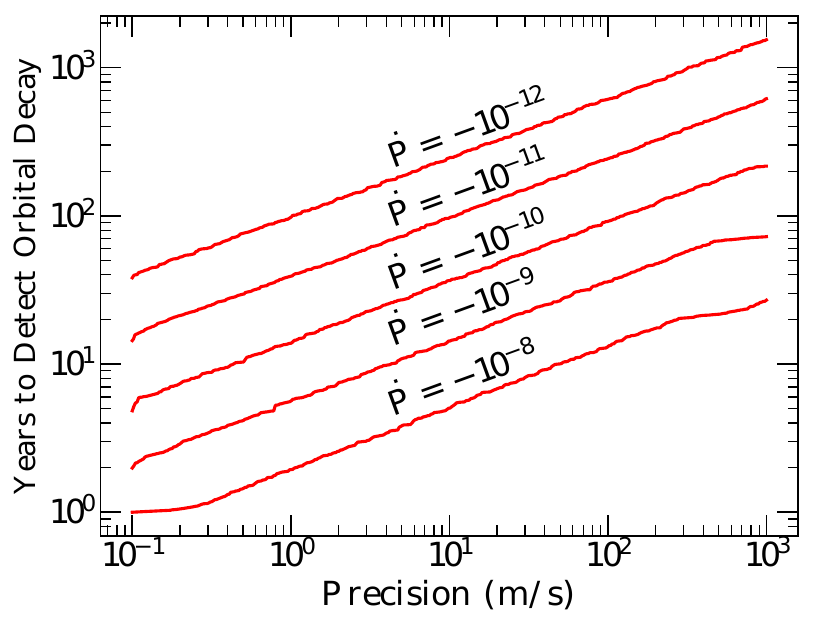}
    \caption{Time required to detect orbital decay in a hot Jupiter for various RV precisions and decay rates.}
    \label{Contours}
\end{figure}

To estimate how long it might take to detect orbital decay in a planet like WASP-12\,b, we carried out the following test. 
For a given orbital decay rate and a given typical uncertainty in RV measurement (precision), we simulated mock RV data spanning some time $T$. We fit the two RV models to the simulated data and calculated the Bayesian Information Criterion (BIC; see Section \ref{GoF}) for each model. If the BIC of the orbital-decay model was smaller by 5 (strong evidence) than the BIC of the constant-period model, then $T$ was considered the time required to detect orbital decay at that chosen orbital decay rate and RV precision. For simplicity, we fixed the following parameters: $e=0$, $i=90^\circ$, $m_p = 1~M_J$, $m_\star= 1~ M_\odot$, $\phi = 0^\circ$, $c = 0 \rm ~m~s^{-1}$, and $P_0 = 1$~day. Figure \ref{Contours} displays the results of these simulations and shows the required duration of RV monitoring for different RV precisions and orbital-decay rates. A planet like WASP-12\,b undergoing orbital decay at its measured rate of $\dot{P} \approx -10^{-9}$ can be detected within about 4~yr with RV precision of $1 \rm ~m~s^{-1}$. Although this level of precision is achievable in principle by a few RV instruments (e.g., HARPS, ESPRESSO; \citealt{2003Msngr.114...20M, 2021A&A...645A..96P}), in practice, stellar jitter will limit the precision achievable for Sun-like stars. As precision in RV measurements improves in the future, it will become easier to detect orbital decay in many hot Jupiters. 


For WASP-12,
WASP-43, WASP-103, and HAT-P-23, we fit the RV measurements that were collated from the literature to the models (Eq.~\ref{constant-model} and Eq.~\ref{decay-model}) using \texttt{emcee}. We kept $P$, $\phi$, and $c$ as free parameters for the constant-period model, while keeping $m_p$, $m_\star$, and $i$ fixed to the most recent values found in the literature. Similarly, we allowed $\dot{P}$, $P_0$, $\phi$, and $c$ to remain as free parameters for the orbital-decay model. Table \ref{radialvelocityfits} lists the best-fit values from the fitting analysis. Table \ref{tab:rv-data} shows all the RV measurements that were collected and used for this analysis.

\begin{table*}
    \caption{Best-Fits Parameters for the Radial-Velocity Analysis}
    \centering
    \begin{tabular}{c|c|c|c|c|c|c|c}
         \hline
         \hline
          & Constant Model & & & Orbital Decay \\
         System & P (days) & $\phi$ & $\mathrm{c}$ (m~s$^{-1}$) & P (days) & $\dot{P}$ & $\phi$ & $\mathrm{c}$ (m~s$^{-1}$) \\
         \hline
         WASP-12 & 1.091428 (2) & 87$^{\circ} \pm$ 2$^{\circ}$ & 20 $\pm$ 1 & 1.0914195 (11) & (-0.32 $\pm$ 0.84) $\times$ $10^{-10}$ & 72$^{\circ} \pm$ 1$^{\circ}$ & 20.9 $\pm$ 0.8 \\
         WASP-43 & 0.813472 (2) & -48$^{\circ} \pm$ 5$^{\circ}$ & 4$\pm$3 & 0.813481 (8) & (-8.2 $\pm$ 6.3) $\times~10^{-10}$ & -35$^{\circ} \pm$ 11$^{\circ}$ & 3 $\pm$ 2 \\
         WASP-103 & 0.925460 (33) & -48$^{\circ} \pm$ 4$^{\circ}$ & 3 $\pm$ 9 & 0.925461 (32) & $(0.3 \pm 7.0) \times 10^{-10}$ & -45$^{\circ} \pm$ 5$^{\circ}$ & 3 $\pm$ 9 \\
         HAT-P-23 & 1.212871 (15) & -48$^{\circ}$  $\pm$ 17$^{\circ}$  & 11 $\pm$ 11 & 1.212889 (15) & (0.7 $\pm$ 7.2) $\times$ $10^{-11}$ & 
         \-28$^{\circ}$ $\pm$ 24$^{\circ}$  & 4 $\pm$ 8 \\ 
         \hline
    \end{tabular}
    \label{radialvelocityfits}
\end{table*}

\begin{table*}
    \caption{Radial-Velocity Data from the Literature}
    \centering
    \begin{tabular}{c|c|c|c|c}
         \hline
         \hline
         System & $\mathrm{BJD}_{\mathrm{TDB}}$ & RV (m~s$^{-1}$) & $\sigma_{\mathrm{RV}}$ (m~s$^{-1}$) & Reference \\
         \hline
         HAT-P-23 & 2454638.10676 & -137.90 & 39.19 & \cite{2011ApJ...742..116B} \\
         HAT-P-23 & 2454674.91479 & 374.64  & 39.85 & \cite{2011ApJ...742..116B} \\
         HAT-P-23 & 2454723.78794 & -209.16 & 40.25 & \cite{2011ApJ...742..116B} \\
         HAT-P-23 & 2454725.85384 & 329.57  & 40.88 & \cite{2011ApJ...742..116B} \\
         HAT-P-23 & 2454726.88051 & 277.98  & 38.9 & \cite{2011ApJ...742..116B} \\
         \hline
    \end{tabular}\\
    {\raggedright \textbf{Note.} The full table is available online. See the Data Availability statement. The barycentric radial velocity was subtracted from the RV measurements.
    \par}
    \label{tab:rv-data}
\end{table*}

\subsection{Goodness-of-Fit Metrics and Joint Fitting} \label{GoF}

To better understand whether one model is favoured over the other, we calculated the BIC, and the reduced chi-squared ($\chi^2$ per degree of freedom = $\chi^2_r$). BIC is defined  as
\begin{equation}
    \mathrm{BIC} = \chi^2 + k \log\,n, \end{equation}
where $n$ is the number of data points and $k$ is the number of free parameters in the model \cite{1978AnSta...6..461S}. The model which has low $\chi^2$ with fewer free parameters is favoured by BIC. With the use of these two statistical metrics, we can make inferences as to which model better explains the data. 

For the systems that have sufficient number of RV measurements (WASP-12, WASP-43, WASP-103, and HAT-P-23), we also performed fitting of a joint transit-RV model by combining data from the transit and RV measurements.In Table \ref{Joint-Fit}, we show our best-fit parameters from the joint analysis of these systems. 

\begin{table*}
    \caption{Best-Fit Parameters for Joint Transit-RV Fitting}
    \centering
    \setlength\tabcolsep{2.5pt}
    \begin{tabular}{c|c|c|c|c|c|c|c|c|c|c}
    \hline
    \hline
          & Constant  & & & & Orbital Decay  \\
         System & $P$ (days) & $t_0$ ($\mathrm{BJD}_{\mathrm{TDB}}$) & $\phi$ & c (m~s$^{-1}$) & $P$ (days) & $t_0$ ($\mathrm{BJD}_{\mathrm{TDB}}$) & $\dot{P}$ &  $\phi$ & c (m /s) \\
         \hline
        WASP-12 & 1.09141892 (4) & 2457010.51203 (1) & 71.2$^{\circ}\pm$0.2$^{\circ}$ & 21.6$\pm$0.6 & 1.09141942 (3) & 2457010.51292 (3) & (-8.76$\pm$0.34)$\times 10^{-10}$ & 62.4$^{\circ}\pm$0.5$^{\circ}$ & 20$\pm$1 \\
        WASP-43 & 0.81347 (14) & 2457423.44971 (4.5) & -40.6$^{\circ} \pm 0.2^{\circ}$  & 6 $\pm$ 1 & 0.81347405 (4) & 2457423.44977 (9) & (-0.52$\pm$ 0.69) $\times 10^{-10}$ & -41$^{\circ} \pm 1^{\circ}$ & 6 $\pm$ 1 \\
        WASP-103 & 0.92554541 (6) & 2457511.94450 (4) & -264$^{\circ} \pm 2^{\circ}$ & -2$\pm$8 & 0.92554 (13) & 2457511.94450 (4) & (-0.12 $\pm$ 1.16) $\times 10^{-10}$ & -265$^{\circ} \pm 3^{\circ}$ & -1$\pm$8 \\
        HAT-P-23 & 1.2128864 (1) & 2456500.57793 (7) & -26$^{\circ} \pm 1^{\circ}$ & 3$\pm$6 & 1.212886 (26) & 2456500.57792 (7) & (4.8.0 $\pm$ 6.6)$\times 10^{-11}$  & -29$^{\circ}\pm$1$^{\circ}$ & 3$\pm$5 \\
        \hline
    \end{tabular}
    \label{Joint-Fit}
\end{table*}

\section{Results} \label{Results}

\subsection{WASP-12\,b}

\begin{figure*}
    \centering
    \includegraphics[width=\textwidth]{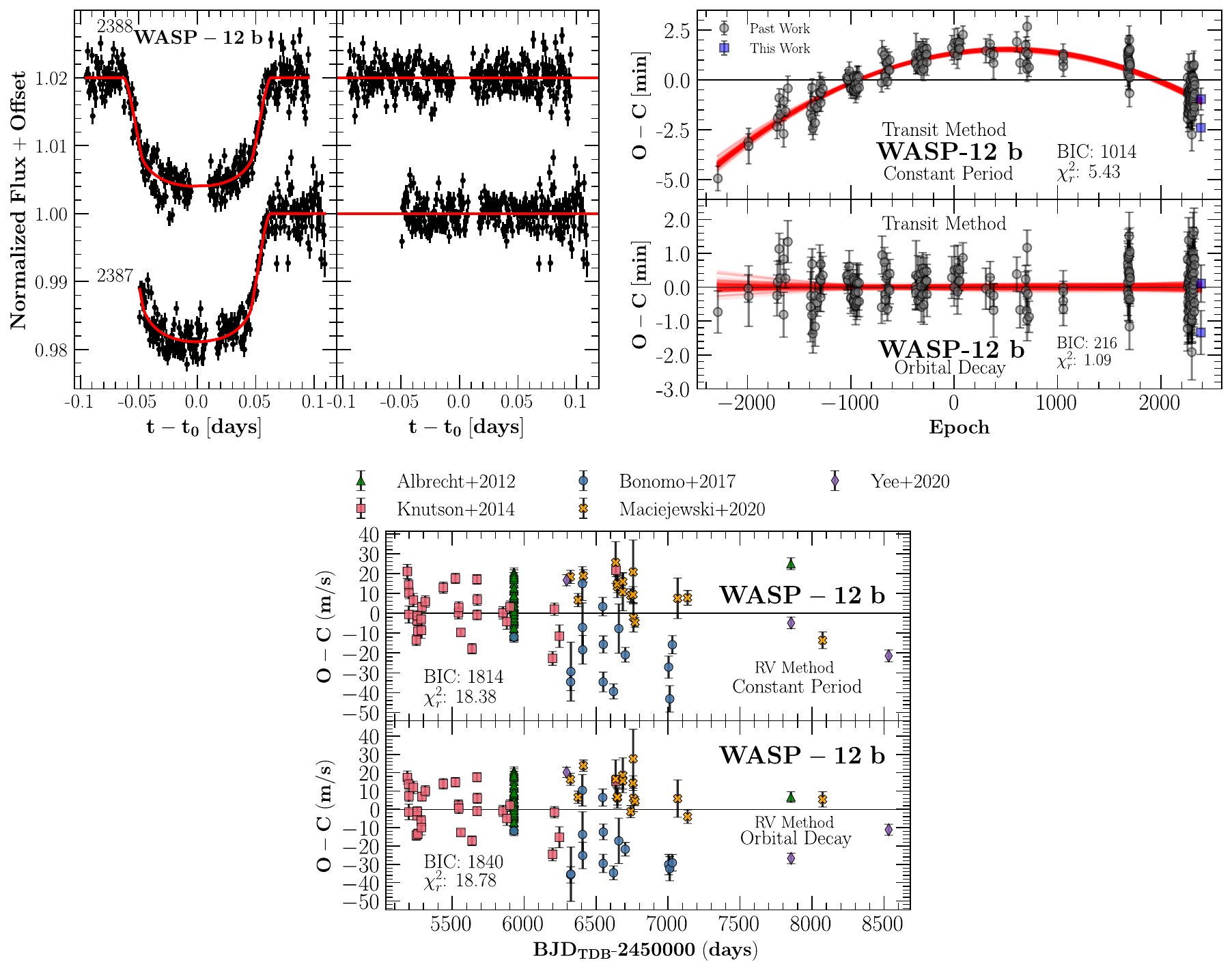}
    \caption{Upper left subplot: The left panel presents the new light curves of WASP-12, with epoch numbers indicated on the left-hand side. The right panel shows the residuals corresponding to each light curve. Upper right subplot: The top panel depicts the transit times subtracted from the best-fit constant-period model. The bottom panel displays transit times subtracted from the best-fit orbital-decay model. We randomly selected 100 posterior samples from the orbital decay model, shown as red curves in both the top and bottom panels. Bottom subplot: The top panel illustrates the residuals from the RV fitting using a constant-period model, while the bottom panel shows the residuals for the RV method under an orbital-decay model.}
    \label{W12-Plot}
\end{figure*}

WASP-12\,b is a 1.5\,$\mathrm{M_J}$ planet that orbits in 1.09 days around a late F-type star of mass 1.4\,$\mathrm{M}_\odot$ \citep{2017AJ....153...78C}. \cite{2016A&A...588L...6M} and \cite{2017AJ....154....4P} first hinted at orbital decay by demonstrating that the orbital period of WASP-12\,b was decreasing, but neither study could differentiate between orbital decay and apsidal precession of the orbit. 
It was not until \cite{2020ApJ...888L...5Y} ruled out apsidal precession that the rate of orbital decay was confirmed to be $-29 \pm 2$~ms~yr$^{-1}$. Figure \ref{W12-Plot} (left panel) shows the new light curves of WASP-12\,b observed here. The right panel displays the transit timing residuals -- also known as ``observed minus calculated'' (O--C) since it is the difference between the observed data and the best-fit model. The bottom panel represents the RV residuals. 

\textbf{Transit:} 
Using our new transit times combined with times from the literature, we determined that the best fit for $\dot{P}$ in the orbital decay model was to $(-9.37 \pm 0.33) \times 10^{-10}$, or equivalently $-29.5 \pm 1.0$~ms~yr$^{-1}$. We use this value of $\dot{P}$, as well as $\mathrm{M_p} / \mathrm{M}_\star$ and $\mathrm{R}_\star /\mathrm{a}$ from \cite{2017AJ....153...78C}, to calculate the tidal quality factor $Q_\star' = (1.6 \pm 0.1) \times 10^{5}$. Our results are consistent with the most recent study of orbital decay for WASP-12\,b, which found $\dot{P} = (-29.81 \pm 0.94)$~ms~yr$^{-1}$ \citep{Wong_2022}. The same authors found $Q_\star' = (1.50 \pm 0.11) \times 10^5$. By fitting for acceleration of orbital decay, we found $\ddot{P} = (-7 \pm 8) \times 10^{-14} \rm ~s^{-1}$. Currently, there is no evidence of accelerated decrease of orbital period, but we can rule out $\lvert \ddot{P} \rvert > 2 \times 10^{-13} \rm ~s^{-1}$ with 95\% confidence. This constraint on $\ddot{P}$ is consistent with the \citet{1966Icar....5..375G} theory of tidal dissipation, which predicts $\ddot{P} \approx -10^{-23} \rm ~s^{-1}$ for WASP-12\,b. Although this is not a strong constraint, continued monitoring of WASP-12\,b will put even more stringent constraints on $\ddot{P}$, thereby testing the validity of the classical equilibrium tidal theory for explaining the orbital decay of hot Jupiters. 

\textbf{Radial Velocity:}
The time stamps on RV measurements from \cite{2017A&A...602A.107B}, which were in $\mathrm{BJD}_{\mathrm{UTC}}$, were converted to $\mathrm{BJD}_{\mathrm{TDB}}$ following the work of \cite{Eastman_2010}. We excluded the RV measurements from \citet{2009ApJ...693.1920H} and \citet{2011MNRAS.413.2500H} because the timing standard was not specified. In all, we used RV measurements from \citet{2012ApJ...757...18A}, \citet{ 2014ApJ...785..126K}, \citet{2017A&A...602A.107B}, \citet{2020ApJ...888L...5Y}, and \citet{2020ApJ...889...54M}. 
The barycentric velocity, which was calculated to be $\gamma=19.1 ~\rm km~s^{-1}$, was subtracted from the RVs provided by \cite{2017A&A...602A.107B} and \cite{2020ApJ...889...54M}. We removed a data point from \citet{ 2014ApJ...785..126K} as this was an outlier in our analysis.

From fitting of the RV measurements alone, we get $\dot{P} =  (-8 \pm 17) \times 10^{-10}$. Currently, the RV measurements of WASP-12\,b are unable to detect orbital decay, even though the transit times show overwhelming evidence of a decreasing orbital period. 
We note that there is an offset between the \citet{2020ApJ...889...54M} and the \citet{2017A&A...602A.107B} datasets. This offset may be due to imperfect subtraction of barycentric velocity, which is unknown in the two papers. We carried the analysis after omitting the two datasets, and found no orbital decay.

\textbf{Goodness-of-Fit Metrics and Joint Fitting:} For the transit method, the constant-period model has $\chi^2_r = 5.43$ and BIC = 1014, whereas the orbital decay model has $\chi^2_r = 1.09$ and BIC = 216. The orbital-decay model is thus decisively favoured over the constant-period model.

Similarly, we examined the goodness-of-fit metrics for the RV fitting. For the constant-period model, we calculated $\chi^2_r = 18.38$ and BIC = 1814. On the other hand, for the orbital-decay model, we found $\chi^2_r = 18.78$ and BIC = 1840. Although the constant-period model has a slightly lower $\chi^2$ compared to the constant-period model, no model is significantly more favourable than the other based on the RV measurements. In fact, both models perform quite poorly in describing the RV data, suggesting  unaccounted systematic errors in the RV measurements. 

In the joint RV-transit fitting, we found 
that the constant-period model has BIC = 1462 and $\chi^2_r = 5.08$. The orbital-decay model had $\dot{P}$ = ($-8.76 \pm 0.34)\times 10^{-10}$, with BIC = 482 and $\chi^2_r = 1.61$. It is noteworthy that the value of $\dot{P}$ for joint fitting nearly agrees with the value found using the transits alone, indicating that the joint fitting is dominated by the transits while the RV measurements contribute little. Overall, the orbital-decay model is decisively favoured over the constant-period model.

\subsection{WASP-43\,b}

\begin{figure*}
    \includegraphics[width=\textwidth]{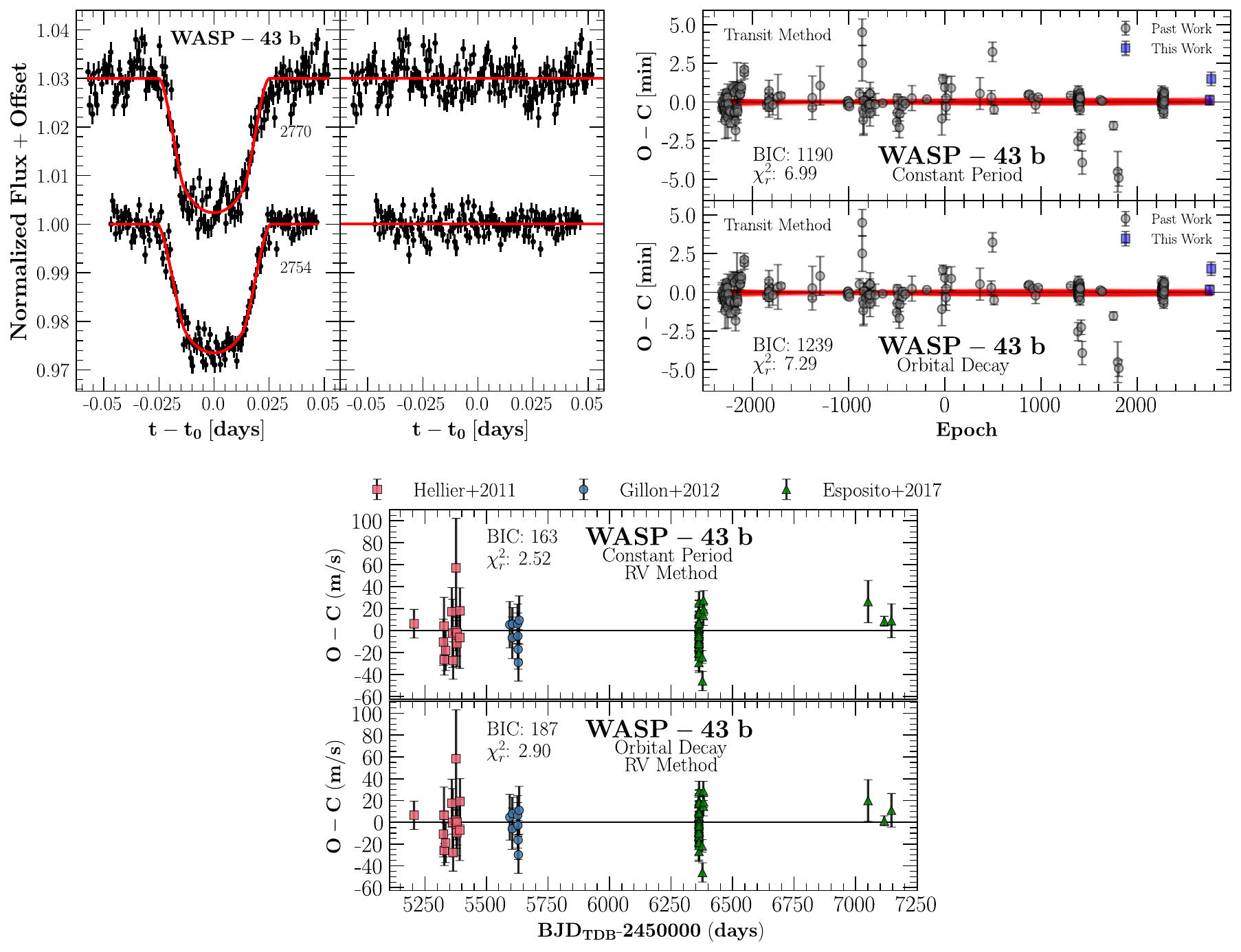}
    \caption{Same as Figure \ref{W12-Plot}, but for WASP-43\,b.}
    \label{W43-plots}
\end{figure*}

Figure \ref{W43-plots} (left panel) shows the new light curves obtained here. The right panel and the bottom panel show the residuals for transit timing and RV, respectively. 

\textbf{Transit}:
WASP-43\,b has a mass of 2.05\,$\mathrm{M_J}$ and orbits a K7V-type, 0.7\,$\mathrm{M}_\odot$ star every 0.81 days \citep{2012A&A...542A...4G, 2011A&A...535L...7H}. \cite{2021AJ....162..210D} found that $\dot{P} = (-1.11 \pm 0.21) \times 10^{-10}$ and $Q_\star' > 4.0 \times 10^{5}$. Using past literature in combination with our transits, we determined that $\dot{P} = (-9 \pm 51) \times 10^{-12}$, or a rate of $-0.2 \pm 1$~ms~yr$^{-1}$. We used values of $\mathrm{M_p}$, $\mathrm{M}_\star$,  $\mathrm{R}_\star$, and $a$ from \cite{2021AJ....162..210D} and $\dot{P}$ from our work to calculate a lower limit of $Q_\star' > 3.9 \times 10^5$. While there is no orbital decay in WASP-43\,b at the moment, we present an improvement of the transit ephemeris to $t(E)= 2457423.44971(4) \mathrm{BJD_{TDB}} + E\times0.81347405(2)$.

\textbf{Radial Velocity}:
RV measurements were collected from \citet{2011A&A...535L...7H}, \citet{2012A&A...542A...4G}, and \citet{2017A&A...601A..53E}. The time stamps on RV measurements by \citet{2011A&A...535L...7H} were converted from $\mathrm{BJD}_{\mathrm{UTC}}$ to $\mathrm{BJD}_{\mathrm{TDB}}$ as described by \cite{Eastman_2010}. Fits to the RV measurements after subtracting the barycentric radial velocity showed $\dot{P} = (-8.2 \pm 6.3) \times 10^{-10}$. The joint RV-transit analysis also indicates no orbital decay in the system, with $\dot{P} = (-0.52 \pm 0.69) \times 10^{-10}$.

\subsection{WASP-103\,b}

\begin{figure*}
    \includegraphics[width=\textwidth]{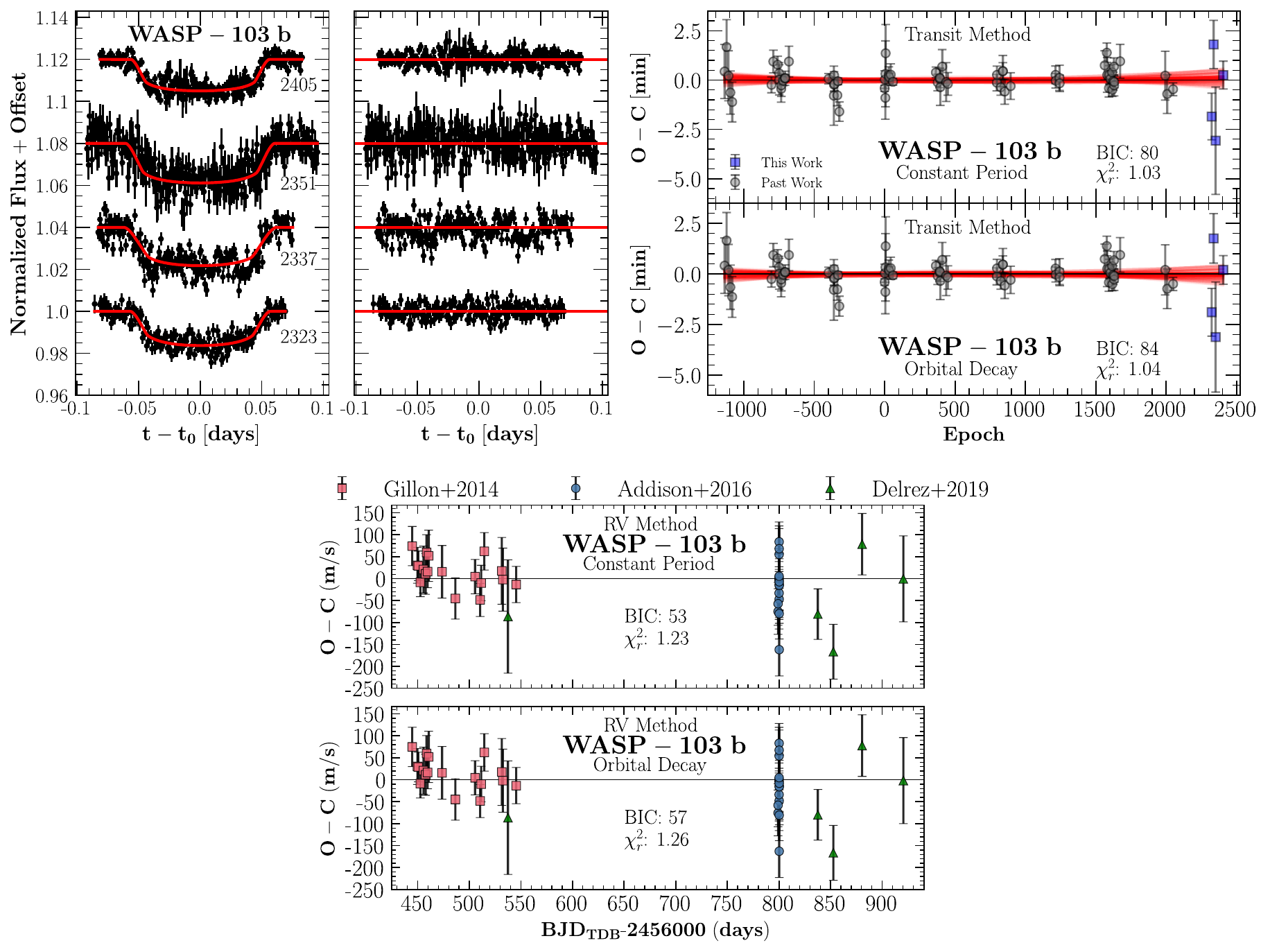}
    \caption{Same as Figure \ref{W12-Plot}, but for WASP-103\,b.}
    \label{W103-PLOTS}
\end{figure*}

Figure \ref{W103-PLOTS} (left panel) shows the new light curves obtained here. The right panel and the bottom panel show the residuals for transit timing and RV, respectively. 

\textbf{Transit}:
WASP-103\,b was discovered in 2014 with a mass of 1.5\,$\mathrm{M_J}$, orbiting an F8 star of 1.22\,$\mathrm{M}_\odot$ every 0.92 days \citep{2014A&A...562L...3G}. \cite{2017MNRAS.472.3871T} searched for transit-timing variations (TTVs) and found none. \cite{2020AJ....159..150P} also studied TTVs of WASP-103\,b, but found the rate of orbital decay to be marginally positive. They also derived a lower-limit value of $Q_\star'$ to be $>(1.1 \pm 0.1) \times 10^{5}$ with $95\%$ confidence. \cite{2022MNRAS.512.2062B} provide the most updated values of $\dot{P}$ and $Q_\star'$ to be $-1.1 \pm 2.3$~ms~yr$^{-1}$ and $>1.4 \times 10^{6}$, respectively.

By combining times from past literature with our transit times, we found $\dot{P} = (2.49 \pm 9.94) \times 10^{-11}$, or a rate of $(-0.78 \pm 3.13)$~ms~yr$^{-1}$. Using the values of $\mathrm{M_p}$, $\mathrm{M}_\star$, and $\mathrm{R}_\star /\mathrm{a}$ from \cite{2022MNRAS.512.2062B}, and our value of $\dot{P}$, we determine that $Q_\star' > 1.2 \times 10^6$. This result is consistent with past work. \cite{2022A&A...667A.127M} report $Q_\star' >3.8\times 10^6$, an even tighter constraint. We refined the constant-period ephemeris to $t(E)=2457511.94449(3)\mathrm{BJD_{TDB}}+E\times0.92554540(3)$.

\textbf{Radial Velocity}:
RV measurements were collected from \cite{2014A&A...562L...3G}, \cite{2016ApJ...823...29A}, and \cite{2018MNRAS.474.2334D}. Time stamps from \cite{2016ApJ...823...29A} were converted from $\mathrm{BJD}_{\mathrm{UTC}}$ to $\mathrm{BJD}_{\mathrm{TDB}}$ as outlined by \cite{Eastman_2010}.  
After subtracting the barycentric radial velocity from the measurements of \cite{2014A&A...562L...3G}, \cite{2016ApJ...823...29A}, and \cite{2018MNRAS.474.2334D}, we determined $\dot{P} = (0.3 \pm 7.0) \times 10^{-10}$, indicating no orbital decay in WASP-103\,b. The joint RV-transit analysis also shows no orbital decay, with $\dot P = (-0.12 \pm 1.16) \times 10^{-10}$. 

\subsection{HAT-P-23\,b}

\begin{figure*}
    \includegraphics[width=\textwidth]{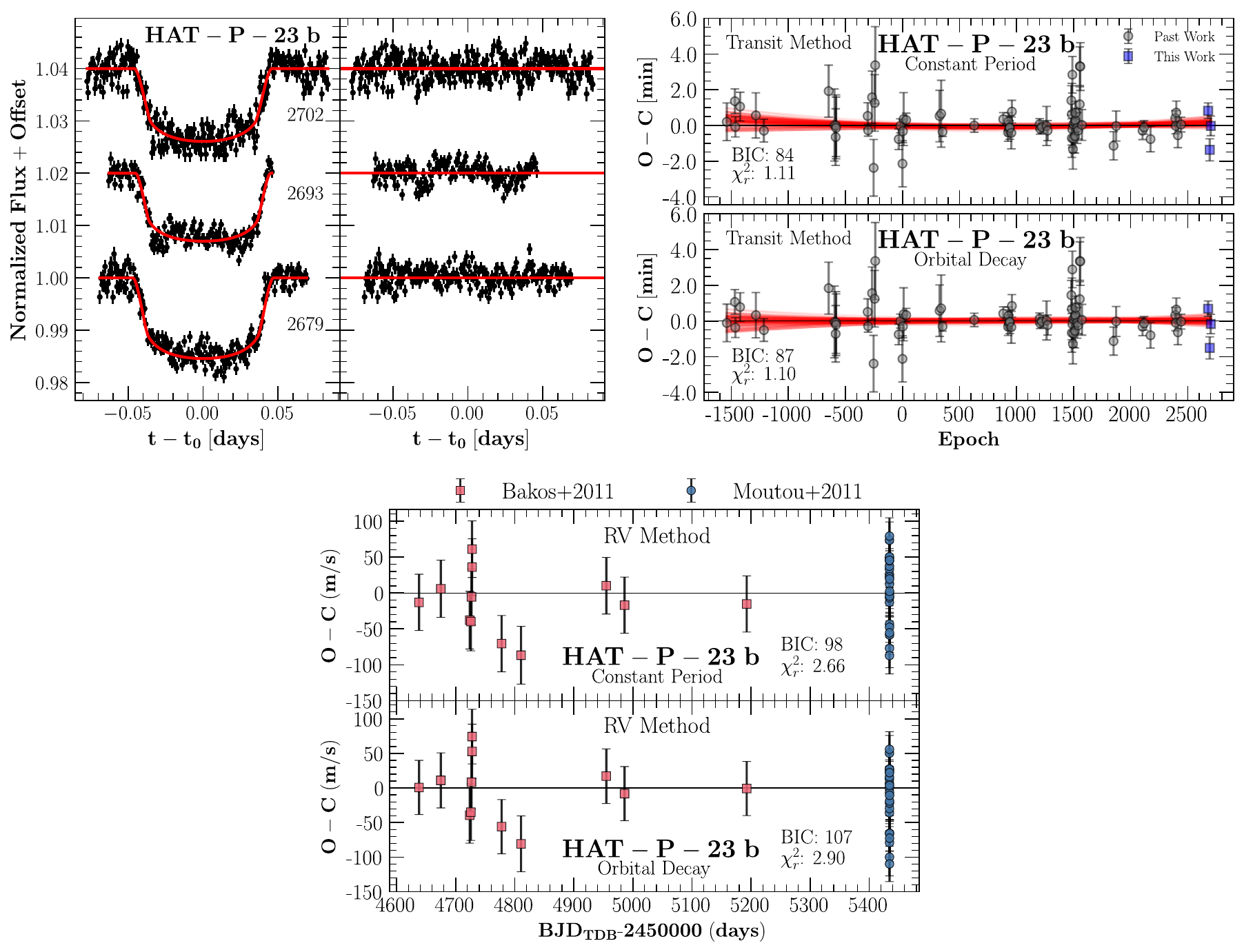}
    \caption{Same as Figure \ref{W12-Plot}, but for HAT-P-23\,b.}
    \label{HAT-PLOTS}
\end{figure*}

Figure \ref{HAT-PLOTS} (left panel) shows new light curves for HAT-P-23b presented here. The right panel displays the timing residuals, and the bottom panel shows the RV residuals.

\textbf{Transit:} HAT-P-23\,b is a  2.1\,$\mathrm{M_J}$ planet that orbits a G0 star of 1.1\,$\mathrm{M_\odot}$ every 1.2 days \citep{2011ApJ...742..116B}. \cite{2018AcA....68..371M} placed a lower limit on $Q_\star'$ with 95$\%$ confidence of $5.6 \times 10^{5}$. It was then revised by \cite{2020AJ....159..150P} to a lower limit of $(6.4 \pm 1.9) \times 10^{5}$. However, the most updated value is from \cite{2022A&A...667A.127M}; they put a constraint of $>(2.76 \pm 0.21)\times 10^6$. Using transit times from past literature and our transit times, we calculated $\dot{P} = (7.42 \pm 5.77) \times 10^{-11}$, or a rate of $2.3 \pm 1.8$~ms~yr$^{-1}$. We used the values of $\mathrm{M_\star}$, $\mathrm{M_P}$, $\mathrm{R_\star}$, and $a$ from \cite{2015A&A...577A..54C} and $\dot{P}$ from this study to calculate $Q_\star' > 9.4 \times 10^5$. Past work from \cite{2018AcA....68..371M} is consistent with this result. In our analysis of HAT-P-23\,b, we exclude two mid-transit times from \cite{2022MNRAS.512.2062B}; the same authors also exclude those two times because of data-quality concerns. With this dataset, we refined the transit ephemeris to $t(E)=2456500.57792(6)\mathrm{BJD_{TDB}} + E \times 1.21288648(4)$.

\textbf{Radial Velocity:}
We used RV data from \cite{2011ApJ...742..116B} and \cite{2011A&A...533A.113M}. The time stamps were converted into $\mathrm{BJD}_{\mathrm{TDB}}$ as outlined by \cite{Eastman_2010}. The barycentric radial velocity was subtracted from the RV measurements of \cite{2011A&A...533A.113M}. 
The best-fit parameter for $\dot{P}$ was determined to be $(0.7 \pm 7.2) \times 10^{-11}$. There is no evidence of orbital decay in HAT-P-23\,b. 

\subsection{KELT-16\,b}

\begin{figure*}
    \includegraphics[width=\textwidth]{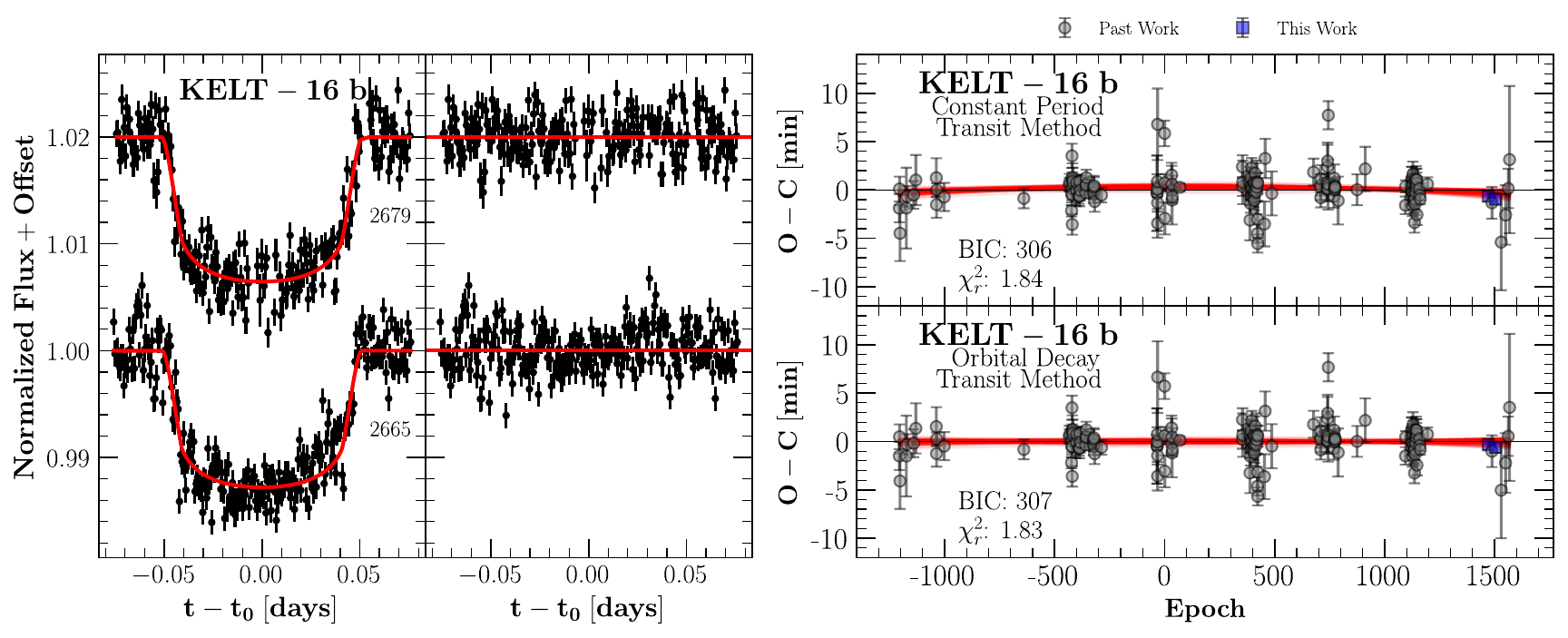}
    \caption{Left subplot: The left panel displays all light curves along with their associated epochs for KELT-16 as presented in this study. The right panel shows the residuals for each corresponding light curve. Right subplot: The top panel presents the O–C residuals for a constant-period model, while the bottom panel shows the residuals for an orbital-decay model. The red curves show 100 random posterior samples of the orbital decay model in both the top and bottom panels.}
    \label{K16-PLOTS}
\end{figure*}

In Figure \ref{K16-PLOTS} (left panel), we showcase the light curves that were collected. The right panel displays the timing residuals.

\textbf{Transit:}
KELT-16\,b is a 2.75\,$\mathrm{M_J}$ planet that orbits an F74 star of 1.2\,$\mathrm{M_\odot}$ every 0.96 days \citep{2017AJ….153...97O}. We find that $\dot{P} =  (-3.94 \pm 3.12) \times 10^{-10}$, or a rate of $-12 \pm 9$~ms~yr$^{-1}$. Using the values of $\mathrm{M_\star}$, $\mathrm{M_P}$, $\mathrm{R_\star}$, and $\mathrm{a}$ from \cite{2022MNRAS.509.1447M} and our value of $\dot{P}$, we determine a lower limit of $Q_\star' > 2.2 \times 10^5$. These values are consistent with the most recent study that attempts to detect orbital decay for this system and finds $\dot{P} = (-2.81 \pm 3.89) \times 10^{-10}$ \citep{2023A&A...669A.124H}. In the same study, the authors gave a lower limit of $Q_\star' > 2.1 \times 10^{5}$. However, \cite{2022A&A...667A.127M} provide an even tighter constraint with $Q_\star' > 3.0\times 10^5$. The improved constant-period ephemeris for KELT-16\,b is $t(E)=2458334.45807 (8)\mathrm{BJD_{TDB}} + E \times 0.9689928 (11)$.

\subsection{WD 1856+534\,b}
\begin{figure*}
    \includegraphics[width=\textwidth]{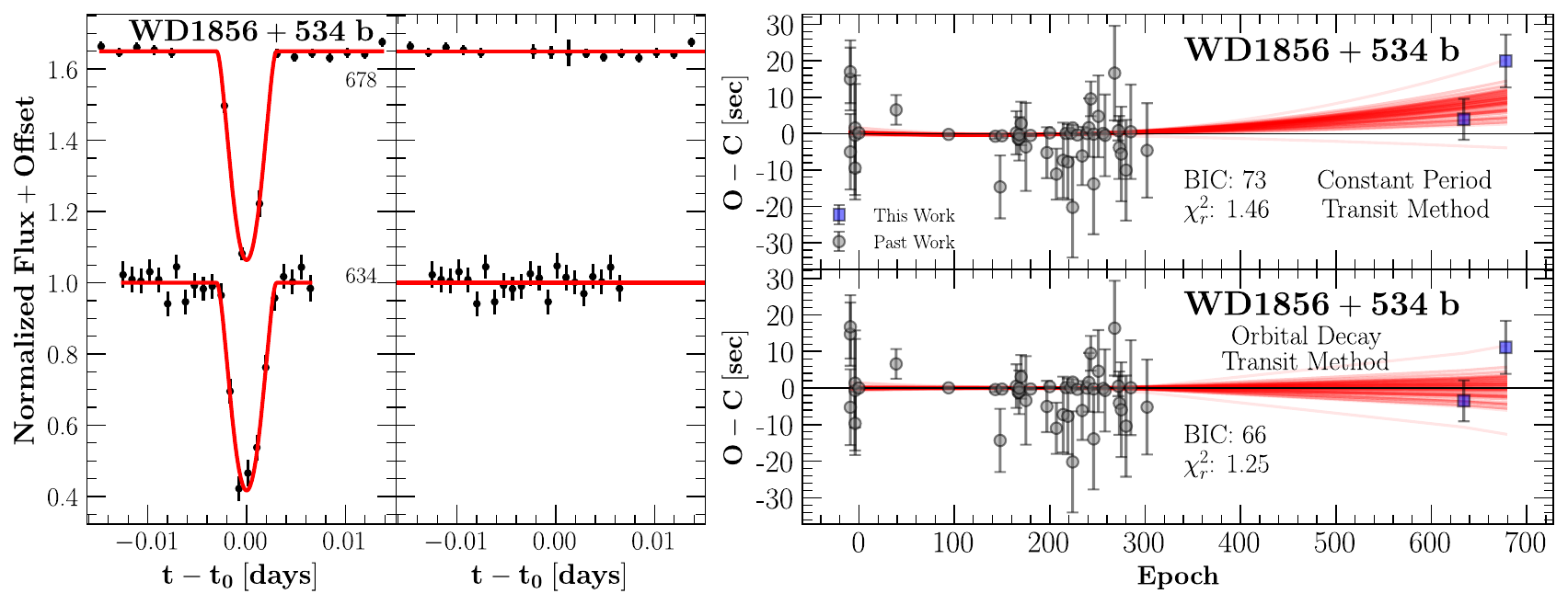}
    \caption{Same as Figure \ref{K16-PLOTS}, but for WD 1856+534\,b.}
    \label{WD-PLOTS}
\end{figure*}

In Figure \ref{WD-PLOTS} (left panel), we display our light curves  collected here. The right panel shows the timing residuals.

\textbf{Transit:}
WD 1856+534\,b was first discovered by \cite{2020Natur.585..363V} in 2020. This planet orbits a white dwarf (WD) star every $\sim 1.40$ days. Because this is a relatively new system, it is not yet well studied. However, \cite{2023MNRAS.521.4679K} looked at TTVs for this system while trying to detect another planet. The authors suggest that there is no evidence for multiple planets in the system, and were able to fit a constant-period model from past transit times. We present new transit times. Additionally, we fitted a constant-period model and an orbital-decay model with past transit times in the literature in addition to our times. We find that the orbital period may be changing, $\dot{P} = (4.98 \pm 1.54) \times 10^{-10}$ (i.e., $15 \pm 4$~ms~yr$^{-1}$), with the implication that the orbit of WD 1856+534\,b is slowly becoming larger instead of shrinking. While statistically, this is a 3$\sigma$ detection of orbital growth, we exercise caution in making any definitive claim because the apparent positive period derivative relies on the last two data points that were gathered for this study. More observations in the future will be used to determine whether the orbit of this system is indeed growing. Additionally, we provide an updated constant-period ephemeris of $t(E)=2458779.375079 (2)\mathrm{BJD_{TDB}}+E\times1.40793923 (1)$.

To quantify $Q_\star'$, we used values of $\mathrm{M_\star}$ and $\mathrm{R_\star}$ from \cite{2021A&A...649A.131A},$a$ from \cite{2020Natur.585..363V}, and our value of $\dot{P}$. The mass of WD 1856+534\,b remains uncertain owing to a lack of high-quality RV measurements, and therefore, we assumed $\mathrm{M_P}$ = 1 $\mathrm{M_J}$. We found a rather unphysical $Q_\star'$ $>5.8 \times 10^{-5}$ for the white dwarf WD 1856+534. Since the interior structure of a white dwarf is drastically different from that of a main-sequence star, the tidal quality factor for a white dwarf is expected to be  $Q_\star' \sim 10^{12}-10^{15}$ \citep{2023ApJ...945L..24B, 1984MNRAS.207..433C, 2010ApJ...713..239W}. The strong discrepancy with our measurement could imply that the orbital evolution of the planet is not due to tides raised by the planet on the white dwarf. Instead, tides raised on the planet by the white dwarf may be important. Alternatively, the putative positive period derivative in this system could also result from apsidal precession of the orbit, light travel time, or the Romer effect.  

\textbf{Goodness-of-Fitness Metrics:}
For the constant-period model we found BIC = 73.4, and BIC = 66.5 for the orbital-decay (growth) model. With $\Delta$BIC = 6.9, the orbital-decay (growth) model is formally favoured over the constant-period model. Thus, we conclude that this is a tentative detection of orbital growth in WD 1856+534\,b. 

\subsection{WTS-2 b}

\begin{figure*}
    \includegraphics[width=\textwidth]{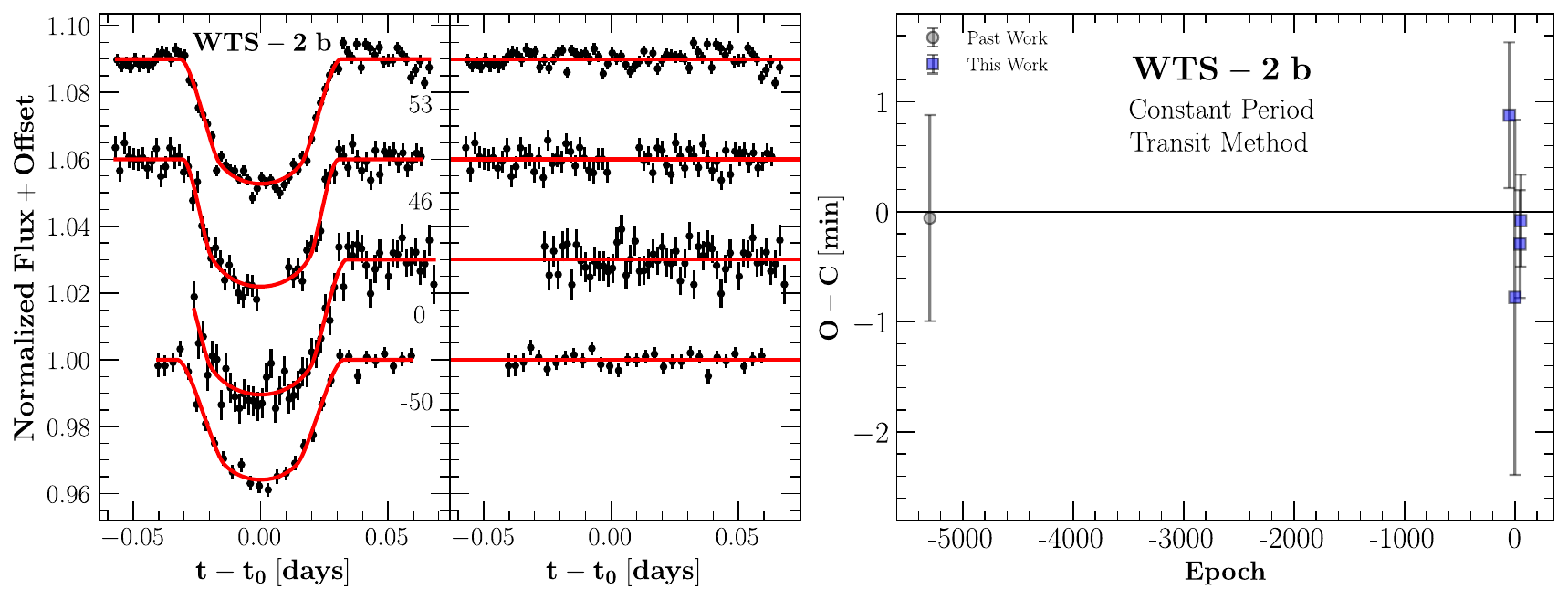}
    \caption{The left panel in the left subplot shows all light curves with their associated epoch for WTS-2 presented here. The right panel shows the residuals of its respective light curve. The right panel shows O--C residuals of a constant-period model.}
    \label{WTS-PLOTS}
\end{figure*}

\begin{figure}
    \centering
    \includegraphics[width=\columnwidth]{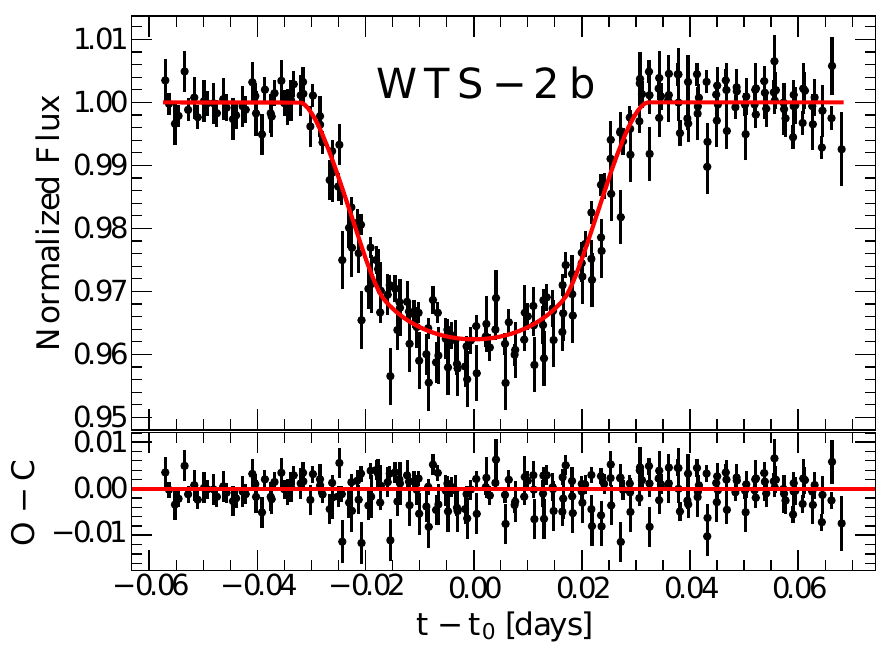}
    \caption{Stacked light curves of WTS-2.} 
    \label{WTS2-stacked}
\end{figure}

We present four new transits shown in the left panel of Figure \ref{WTS-PLOTS}; the right panel shows the O--C for a constant-period model.

\textbf{Transit:}
WTS-2\,b -- first discovered by \cite{2014MNRAS.440.1470B} -- is a 1.12\,$\mathrm{M_J}$ planet that orbits a K3V star of 0.8\,$\mathrm{M_\odot}$ every 1.01 days. Because this is an understudied system, further observations are required to determine whether there is orbital decay in the system. It is important to highlight that we have significantly improved the precision of the orbital period of this system by a factor of $\sim 5$ compared to the discovery paper. We determined an improved constant-period ephemeris to be $t(E)=2459714.91497 (19) \mathrm{BJD_{TDB}} + E \times 1.01870539 (12)$.

Since there are no other transit data available for this system after \cite{2014MNRAS.440.1470B}, we present newly updated transit parameters $R_p / R_\star$, $b$, and $R_\star/ a$. This was done by stacking the four light curves, aligned by their mid-transit times. Subsequently, we use the \cite{2002ApJ...580L.171M} model for transits and fit it with \texttt{emcee} to the stacked light curve. We find that $R_p / R_\star = 0.1849 \pm 0.0028$, $b = 0.611 \pm 0.045$, and $R_\star/ a =  0.1924 \pm 0.0064$. These values are consistent with those from the discovery paper. We also note that there is an improved measurement for $R_\star/a$ by a factor of $\sim 2$, but the other values have uncertainties similar to those in the discovery paper. Figure \ref{WTS2-stacked} shows the stacked light curves with the best-fit model that is used to determine the most updated transit parameters. 

\section{Conclusions} \label{Conclusions}

In this work, we have investigated the planetary systems of WASP-12, WASP-43, WASP-103, HAT-P-23, KELT-16, WD 1856+534, and WTS-2. A total of 19 transit times were observed with the 1~m Nickel telescope at Lick Observatory.
From each transit, we calculated the mid-transit times and also gathered mid-transit times from past literature. Combining both past work and our own current data, we were able to fit two models to the transit timing data: one of constant orbital period and another of orbital decay. 

Our analysis shows that with the exception of WASP-12\,b, none of the observed planetary systems revealed orbital decay. We attempted to  detect acceleration of orbital decay in the WASP-12 system and found that $\ddot{P} = (-7 \pm 8) \times 10^{-14}~ {\rm s}^{-1}$, implying no significant acceleration of decay. The expected acceleration from the classical equilibrium tidal theory, on the other hand, was calculated to be $\ddot P \approx -10^{-23}~ {\rm s}^{-1}$. In the case of WD 1856+534\,b, our data suggest tentatively that this planet is undergoing orbital growth.For the remaining planetary systems that were found to have unchanging orbital period, we provided lower limits on $Q_\star'$ with 95$\%$ confidence. 

Since WTS-2 has been observed only once at the time of writing this paper, we provided updated transit parameters and transit ephemeris. We see that these values are in agreement with those in the discovery paper \cite{2014MNRAS.440.1470B}. Additionally, we improved the precision of the orbital period of WTS-2 by a factor of $\sim 5$ and of $R_\star / a$ by a factor of $\sim 2$.

We also investigated the possibility of detecting orbital decay using RV  data. Our analysis examines systems that have more than 10 RV data points spanning over 2~yr. This narrows our search down to the planetary systems of WASP-12, WASP-43, WASP-103, and HAT-P-23. Owing to the large uncertainties in our best fits for this method, we cannot say with certainty whether the exoplanets are undergoing orbital decay. More precise RV data and longer monitoring duration are needed to achieve meaningful results with this method. Currently, we show the transit method outperforming the RV method. Despite the RV method not providing a tighter constraint on $\dot{P}$, it may be improved by joint fitting with the transit method. With joint fitting, we see smaller uncertainties in $\dot{P}$; however, they are not small enough to rule out a constant-period or an orbital-decay model. Although we do not anticipate that observers will obtain expensive new RV measurements specifically for detecting orbital decay in exoplanets, over time the expanding database of RVs could make it possible to detect period changes in many systems. 

The prospective future of this study would ideally include a longer baseline for both methods of detecting orbital decay. Such studies might detect acceleration of orbital decay in the WASP-12\,b system. Based on our tentative results on WD 1856+543, this could be the first system found to undergo orbital growth. 

\section*{Acknowledgements} 
Research at UC Berkeley is conducted on the territory of Huichin, the ancestral and unceded land of the Chochenyo-speaking Ohlone people, the successors of the sovereign Verona Band of Alameda County. Observations with the 1~m Nickel telescope at Lick Observatory were conducted on the stolen land of the Ohlone (Costanoans), Tamyen, and Muwekma Ohlone tribes. 

We thank Tom Esposito for helpful insights, comments, and discussion. The staff at Lick Observatory provided help and guidance while using the Nickel telescope. We thank our reviewer for their valuable comments on this manuscript.

E.A. thanks the CalNERDS and UC~LEADS programs for mentoring and funding for his work. J.D.S. acknowledges support from the Fannie \& John Hertz Foundation. A.V.F.'s research group at UC Berkeley acknowledges financial assistance from the Christopher R. Redlich Fund, Sunil Nagaraj, Landon Noll, Sandy Otellini, Gary and Cynthia Bengier, Clark and Sharon Winslow, Alan Eustace, William Draper, Timothy and Melissa Draper, Briggs and Kathleen Wood, Sanford Robertson, and numerous other donors. Research at Lick Observatory is partially supported by a generous gift from Google.

\section*{Data Availability}
Data for all observed light curves, transit timing, and RV analysis can be found at \href{https://github.com/efrain-alvarado-iii/Tidal-Orbital-Decay-Data}{https://github.com/efrain-alvarado-iii/Tidal-Orbital-Decay-Data}.



\bibliographystyle{mnras}
\bibliography{citation} 




\appendix




\bsp	
\label{lastpage}
\end{document}